# Spatial and temporal probes in inhomogeneous systems: Theory and experiment

Dragan Mihailovic

"Jozef Stefan" Institute, Jamova 39, 1000 Ljubljana, Slovenia,

Tel. +386 1 477 3388, Fax: +386 1 425 1077

dragan.mihailovic@ijs.si

## *Abstract*

The experimental and theoretical challenges posed by the study of dynamically inhomogeneous systems are outlined in the context of cuprates and other oxides. Considering the pitfalls in the single-component approach to the analysis of inhomogeneous systems, the effect of either temporal or spatial averaging by different experiments is discussed. A group-theoretical symmetry analysis of the observed inhomogeneities in real space and *k*-space observed in the cuprates is shown to lead to a quantitatively verifiable description of the inhomogeneous state, comprising of bound singlet pairs in the ground state and unbound fermions in the excited state. The predicted symmetry breaking associated with pairing is shown to be verifiable experimentally.

## *Introduction*

Dynamic mesoscopic inhomogeneities seem to be a ubiquitous feature of "complex matter". The origin of these inhomogeneities may be anything from the formation of local superconducting pairs in high-$T_c$ superconductors to migrating photoexcited charges on DNA molecules or proteins. The experimental challenge of the last decade has been to invent and perfect new techniques for the investigation of dynamic inhomogeneities on time-scales sufficiently short to freeze the motion of the relevant excitations and give information on the microscopic origin of the dominant interactions leading to the observed complexity.

Theoretical progress in accurately describing inhomogeneities occurring on various length scales depends crucially on high-quality experimental data and a detailed microscopic understanding of the relevant interactions. This challenge is gradually being met by the development of new techniques such as time-resolved X-ray diffraction and the extension of various established techniques to the study of inhomogeneous systems.

Cuprate superconductors is a good case example, where the importance of inhomogeneity was not obvious. Although very soon after the discovery of superconductivity in cuprates, the possible existence of charge inhomogeneity was recognized by Gorkov and Sokol[1] and possibly in the form of spin stripes by Zannen and Gunnarson[2] and others, it has taken quite some time to show experimentally that these systems are indeed inhomogeneous and that these inhomogeneities are relevant[3].

Most experimental techniques discussed in this volume involve some kind of averaging, either spatial or temporal. Not surprisingly, the conclusions reached on the basis of different techniques are sometimes in direct and fundamental disagreement. The origin of these disagreements is believed to be in the interpretation of data, which has been averaged in different ways. As one example, time-resolved optical techniques, which have been discussed in this section give accurate time-domain information on lifetimes and energies of elementary excitations in cuprate superconductors, but involve spatial averaging, and cannot give any direct information on nano-scale spatial structure. However, they show no evidence for a strongly anisotropic superconducting gap (or pseudogap). This is inconsistent with the interpretation of some ARPES experiments, which appear to show an anisotropic gap structure. On the other hand, spectroscopies such as ARPES don't give any nano-scale information either. When interpreted in terms of single-particle spectral functions, the results are clearly not satisfactory for inhomogeneous systems. Interpretation of ARPES and tunneling in terms of a multicomponent picture is very model-dependent and hence more than one interpretation is possible[4]. It is apparent that at present no single experiment can give both spatial and temporal details at once and in order to properly understand nano-scale inhomogeneous systems, a large number of experiments need to be considered together.

The aim here is to discuss a few general issues concerning the study of nanoscale inhomogeneous systems with specific reference to the cuprates. We focus on some specific experiments which are believed to give relevant information on the dynamic structure of these materials. We start by considering the system to be inhomogeneous from the start, and identify experiments which give reliable indications of the symmetry of the dominant interaction. The implications defined by a model interaction which takes these symmetries into account are examined. Experiments which can specifically test the model's predictions are given particular attention.

## *Identification of the length-scale of inhomogeneities.*

The experimental evidence for inhomogeneity comes from many different experiments, ranging from local probe techniques such as XAFS, which give information on near-instantaneous positions of nearest neighbors, to STM which probes the surface charge inhomogeneity. Probably the first experiments which unambiguously reported the presence of an inhomogeneous electronic structure were picosecond Raman experiments which showed the presence of localized states in metallic YBCO[5]. Subsequent time-resolved experiments showed the presence of different intrinsic relaxation times, which confirmed the co-exisence of localised and itinerant states[6]. More recent STM experiments of DeLozanne et al and others show the surface of $YBa_2Cu_3O_{7-\delta}$ [7] and $Bi_2Sr_2CaCu_2O_8$[8] superconductors to be very inhomogeneous, with the length-scale of charge inhomogeneities being of the order of ~2-3 nm. Whether these inhomogeneities are also present in the bulk has to be deduced from other experiments. Even more importantly, the question of whether the inhomogeneity in the bulk is fluctuating or is also static, as on the surface remains to be answered. For some models of superconductivity (e.g. Bose-Einstein condensation) it is essential that the pairs are mobile[9], and this is an important point to determine experimentally.

Just as in real space, nanoscale inhomogeneities are evident also in *k*-space. Inelastic neutron scattering experiments on phonons in hole-doped $YBa_2Cu_3O_{7-\delta}$[10], $La_{2-x}Sr_xCuO_4$[11] and inelastic X-ray scattering in electron-doped $Nd_{1.86}Ce_{0.14}CuO_{4+\delta}$[12] all show an anomaly close to the middle of the Brillouin zone. The wavevector of the anomaly corresponds to $k_0 \sim \pi/l_0$, where $l_0$ is the length-scale of the inhomogeneity is the same as the length scale of inhomogeneity observed in STM images[7,8]. The anomaly appears in a range of wavevectors $\Delta k$ corresponding to the spread in $l_0$.

In contrast, perfectly ordered charge stripes are expected to appear over a much narrower range of *k* than the nano-scale objects seen by STM. Static stripes should give rise to a clearly observable zone folding, but this is not observed. We conclude that the INS and X ray scattering experiments speak for the existence of bulk inhomogeneities which are very similar much like the ones observed on the surface by STM. An analysis performed by Egami et al [11] on $La_{2-x}Sr_xCuO_4$ and $YBa_2Cu_3O_{7-\delta}$ [10] shows that the size of the objects responsible is no larger than a few unit cells.

An anomaly in the electronic dispersion spectra of $La_{2-x}Sr_xCuO_4$, $Bi_2Ca_2CuO_4$, $Bi_2Sr_2CaCu_2O_8$ and, $Nd_{2-x}Ce_xCuO_4$. measured by ARPES[13] is observed at precisely the same point $k_0$ near the middle of the Brillouin zone as the anomaly in the phonon spectrum. A plot of the electronic dispersion from ARPES superimposed on the phonon dispersion spectrum measured by INS is plotted for the case of $La_{2-x}Sr_xCuO_4$ in Figure 1.

Indeed, irrespective of how the interaction in Figure 1 is described, the point in *k*-space where the two excitations cross defines the length-scale of the inhomogeneities occuring in $La_{2-x}Sr_xCuO_4$, $Bi_2Ca_2CuO_4$, $Bi_2Sr_2CaCu_2O_8$, $YBa_2Cu_3O_{7-\delta}$, $Nd_{2-x}Ce_xCuO_4$ and probably other cuprates as well.

Inhomogeneities are expected to be observed also in other oxides and anomalies in the INS are also expected to be observed in non-superconducting materials such as $La_2NiO_4$. However, the details (e.g. temperature-dependence, characteristic size, and general morphology of the nanoscale structure) are expected to be different, and still remain to be determined by a concise comparison with superconducting cuprates. The existence of an anomaly in $La_2NiO_4$ or La$_2$MnO$_4$ implies the existence of charge inhomogeneity, but does not necessarily imply the existence of pairs for example.

## *Experiments which identify the energy scale of the dominant interactions*

The fact that the anomaly appears at the same wavevector in the electronic dispersion and the phonon spectrum is a relatively unambiguous indication that the *e-p* interaction is responsible. Spin-related anomalies in cuprates occur on lower energy scales[14], so we can infer with reasonable certainty that the dominant driving force for the inhomogeneity is phononic in origin.
The energy scale of the anomalies in INS and ARPES discussed above is the same as the maximum magnitude of the pseudogap in the equivalent doping range, as determined by single particle tunneling[15], QP recombination[6], ARPES EDCs[16] and other techniques which measure the charge excitation spectrum. As an example, in Figure 2, we compare the "pseudogap" magnitude determined by QP recombination[6] with the anomaly observed neutron data[10] (shaded area). Here it is worth mentioning that magnetic field measurements show that the field which closes the pseudogap is equal to the Zeeman energy, indicating that pseudogap is indeed the energy between a singlet (pair) ground state and unpaired spin 1/2 excitations[17].

The lifetime $\tau$ of the mixed excitation at $k_0$ deduced from the energy width of the INS peaks[10] $\Delta E \sim 4$ -5 meV (h/$\pi c\tau \sim$ 300 fs) is virtually identical with the superconducting pair recombination lifetime of $\tau_R \sim$ 300 fs. The pair recombination time $\tau_R$ can be measured directly by femtosecond optical techniques using time-resolved pump-probe excited state absorption[6] which probes the QP lifetime or with THz radiation probes[18], which directly measure the condensate recovery as a function of time after de-pairing by a 70 femtosecond optical pulse.

Together, the implication is that the excitation is one and the same: it has the same energy, the same lifetime and the correct temperature dependence (as we shall see later).

## *The symmetry of the interactions and some consequences.*

The point in the BZ which identifies the *e-p* interaction also uniquely identifies the symmetry of the lattice distortions which it causes. A concise group theory analysis shows that an interaction along the ($\xi$,0,0) direction can cause a reduction in point group symmetry within the volume defined by $l_0$. The inhomogeneous state in YBCO consists of objects whose point group is C$_{2v}$, while the surroundings have a higher symmetry (D$_{2h}$).

Knowing the symmetry of the interaction enables one to write a mesoscopic Hamiltonian $H_{MJT}$ with the correct symmetry properties[19] which acts on a length scale defined by $l_0$. This has been done so far for La$_{2-x}$Sr$_x$CuO$_4$ and YBa$_2$Cu$_3$O$_7$. The main conclusion from this analysis is that non-degenerate electronic levels coupling to phonons and spins can only give rise to a symmetric (*s*-wave) deformation, while coupling to doubly degenerate electronic states can give rise to a *d*-wave like interaction in addition to the symmetric one. This new interaction acts on a scale given by the length-scale of the inhomogeneities $l_0$, and may be viewed as a mesoscopic Jahn-Teller effect (to distinguish it from the more standard single-ion JT effect, or band JT effect which involves extended states).

The state which $H_{MJT}$ implies is a dynamically inhomogeneous one, where the distorted and undistorted regions coexist (Fig. 3), and have different energies. The energy difference separating the two states $\Delta$ is associated with the pseudogap. Whether or not electrons (or holes) are paired within these objects can be determined by experiments which measure the spin susceptibility. Pairing would imply the existence of singlets in the ground state and unpaired spins in the excited state. An analysis of NMR data shows this to be very good description of the temperature dependence of the Knight shift[20] (see Fig. 4) as well as static susceptibility[21]. In addition to paired states, the interaction allows the formation of larger objects (stripes), as well as single polarons. The presence of very long stripes in cuprates has so far not been unambiguously reported, while the presence of single isolated polarons would give rise to a substantial Curie susceptibility. Although a small Curie contribution cannot be excluded, fits to susceptibility data[21,22] so far imply that the entire system can be approximated very well by a two-level system with spin singlets in the ground state (due to pairs) and single fermions with spin 1/2 as quasiparticles (see Fig. 4 and ref.20). Thus, we deduce that the most relevant solution to the interaction $H_{MJT}$ is the pair state with dimensions of $l_0 \sim$ 1 - 3 nm[19], nevertheless clustering of these regions is still expected to be important due to elastic forces[23].

## *Discussion: The consequences of the interaction. Symmetry breaking and properties of a 2-component inhomogeneous system.*

Assuming, for simplicity, that the mesoscopically inhomogeneous system has only two components, and that the state can be described by a two-level system of bosons (in the distorted regions caused by interaction $H_{MJT}$) and the surrounding fermions. To test this approximation, we consider the temperature dependence for such a system. This has been calculated previously to describe different experiments[20,21,22,24]. The temperature dependence for the fermion and boson populations is shown in Figure 5. The energy scale is given by the eigenvalue of $H_{MJT}$. Importantly, this temperature dependence shown in Figure 5 describes the *T*-dependence of the anomaly in INS and other spectroscopies, such as XAFS very well (see Figure 5), justifying the simplification of a 2-level system. It is very important to realize that in general, the 2-level system description will not be sufficient when discussing detailed analyses of spectroscopies such as tunneling, ARPES or infrared spectroscopy, which measure the total density of states, or for that matter transport properties. Here, additional localized states, which are not considered by the two-level system, are expected to be important. (Although they may not be particularly relevant for superconductivity.)

### Dynamic probes

EXAFS and PDF probe the local structure on time-scales of the order of $10^{-15}$-$10^{-12}$ s. They effectively freeze the motion of excitations whose energy scales are less than ~500 meV in the case of XAFS, and perhaps a few tens of meV in the case of neutron PDF. Significantly, XAFS structure snapshots are on a timescale, which is faster than the pair recombination time. This means that it can be used to freeze the structure of lattice distortions within local pairs, which occur on an energy scale of the pseudogap (< 100 meV), hence the techniques have been very important in pointing out the presence of inhomogeneities in cuprates and other oxides[25]. XAFS and neutron diffraction PDF data have not yet been analyzed in a way, which would be consistent with the mesoscopic Jahn-Teller deformation discussed above, and would provide a good test of the proposed MJT model.

In this context it is worth pointing out the longstanding puzzle which concerns the fact that STM measurements observe a superconducting gap and a psudo-gap which remains above $T_c$, in spite of the fact that the inhomogeneities are completely static. One possibility is that the surface presents a boundary condition, which pins the inhomogeneities such that they remain fixed on the surface, while the bilk is dynamically changing. Another possibility is that in fact that the bulk has the same nano-scale static inhomogeneity. This would imply that static inhomogeneity (as opposed to long-range ordered stripes) is not detrimental to superconductivity. As already mentioned, such a picture would pose problems for the Bose condensation[9] picture, or superconductivity via stripes[26] and may be difficult to reconcile with some experimental observations, which strongly suggest that the bulk inhomogeneities are dynamic[3]. It should be mentioned here that if pairs can tunnel between inhomogeneities[27,28], it is not particularly important whether the inhomogeneities are dynamic or static, provided the inhomogeneities exist on a timescale which is longer than the pair tunneling time.

### Evidence for loss of inversion symmetry.

According to group theoretical analysis for finite *k*, the group of *k* along the (ξ,0,0) direction of the BZ is $C_{2v}$, and does not contain inversion symmetry (the direction (ξ,0,0) is named Δ in YBCO and Σ in $La_{2-x}Sr_xCuO_4$). As a consequence we expect to observe manifestations of symmetry breaking in spectroscopies, which are sensitive to the presence of a center of inversion, or the presence of a spontaneous polarization. Indeed there is ample evidence for symmetry breaking below the pseudogap temperature, both in Raman[29] and infrared experiments[30]. There is also a large amount of evidence for the existence of a spontaneous polarization in cuprates over a wide range of doping[31]. Such specific (exotic) experimental results are often ignored, in spite of the fact that they may provide the key proof for an accurate theoretical description of the problem. Experiments showing evidence for spatial or time-reversal symmetry breaking such as forbidden phonon Raman and infrared spectra, pyroelectricity, piezoelectricity and magnetic-field induced symmetry-breaking effects give crucial additional information supplementing the results of more standard spectroscopies.

## *Conclusions*

It is abundantly clear that the analysis of a nano-scale inhomogeneous systems cannot be made solely on the basis of averaging methods, either spatial or temporal. On the other hand, it is also clear that for a complete interpretation of complexity in such systems it is necessary to consider more than just one or two experiments, but rather a large number together. Theoretical modeling must be rigorously tested by quantitative examination of many experiments together, but the most important part in the verification of any model are specific predictions, which can be tested by non-routine experiments.

The mesoscopic Jahn-Teller interaction that has been discussed here, is based on concise group-theoretical arguments and has the correct symmetry properties to describe the interaction at finite wavevector $k$. It has been tested on a number of crucial experiments, and has so far been successful in describing the appearance of a new inhomogeneous state with broken spatial symmetry below the pseudogap temperature. However, many more experiments remain to be examined before it can be deemed a successful candidate for a theory of high-temperature superconductivity. Although one should expect that a detailed examination of other experiments which we haven't considered here will reveal additional complexity, (which appears to be an unavoidable feature of inhomogeneous systems) it seems that the two-level system describes the salient features very well. Thus, in cuprates the emerging picture is one in which the scale of the inhomogeneity is of the order of the coherence length, where pairing is coincident with the formation of an inhomogeneous state. Superconductivity can then be shown to follow, even in the case where the inhomogeneities are static.

I wish like to thank Viktor V.Kabanov, S.Conradson, S.Billinge, J.C.Phillips and T.Egami for helpful comments and discussions.

## *Figure captions*

Figure 1. The anomaly in the ARPES electronic dispersion curves and phonon anomaly in inelastic neutron scattering along the $(\xi,0,0)$ direction in the Brillouin zone. The crossing point defines the anomaly in both spectroscopies. The straight line indicates the expected electronic dispersion in the absence of the coupling to phonons. The data are from references 10 and 13.

Figure 2. The "pseudogap" magnitude as a function of doping in YBCO from time-resolved measurements of quasiparticle relaxation[6] and from the anomaly appearing in the INS dispersion[10] and ARPES[13]. The data are consistent with other experimental techniques which measure the charge gap, such as tunneling for example[15].

Figure 3. The inhomogeneous state suggested by $H_{MJT}$. The picture is similar to the STM image of the surface of $Bi_2Sr_2CaCu_2O_8$ in ref. 8.

Figure 4. The NMR Knight shift measured on $^{17}O$ and $^{63}Cu$ and fit using the two-level system model[20] in Y:124. The value of the pseudogap obtained from the fits are also discussed in ref.20. The data are from Curro et al[32].

Figure 5. a) The populations of Fermions and Bosons predicted on the basis of the 2-level system description of the inhomogeneous state. The value of the "pseudogap" used to calculate the model is $\Delta_p = 44$ meV. b) The temperature dependence of the intensity of the anomaly seen in INS [10] using the same value of $\Delta_p$. c) The XAFS in $La_{2-x}Sr_xCuO_4$ shows a similar 2-level system temperature dependence[33].

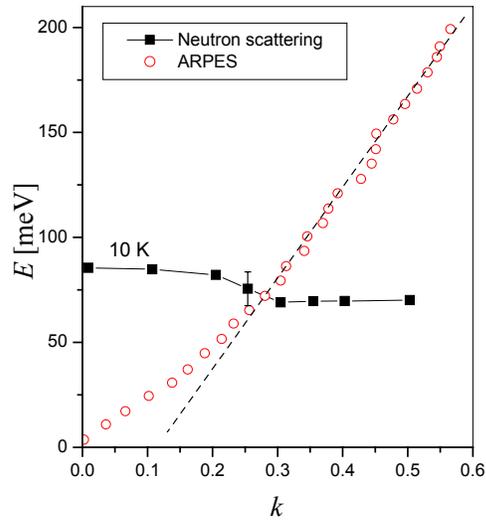

Figure 1

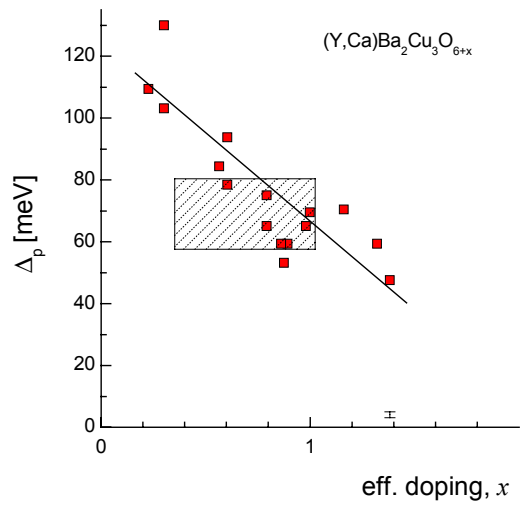

Figure 2

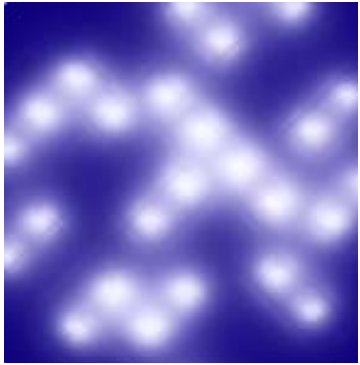

Figure 3

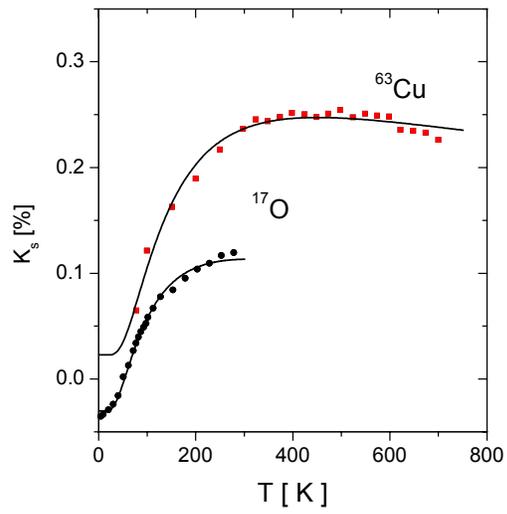

Figure 4

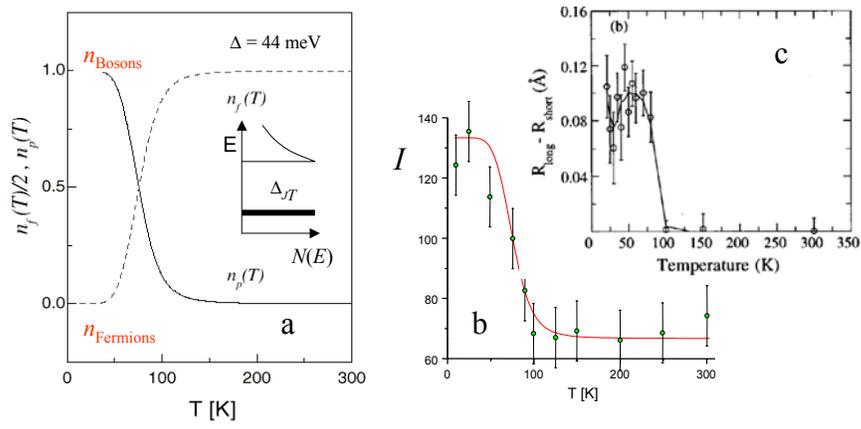

Figure 5